\documentclass[article,nofootinbib,superscriptaddress,notitlepage]{revtex4-1}
\usepackage{graphicx}
\usepackage{subfigure}
\usepackage{slashed}
\usepackage{simplewick}
\usepackage{feynmf}
\usepackage{color}
\usepackage{rotating}

\newcommand{\PRE}[1]{{#1}} % Use if preprint style

\newcommand{\be}{\begin{equation}}
\newcommand{\ee}{\end{equation}}
\newcommand{\bea}{\begin{eqnarray}}
\newcommand{\eea}{\end{eqnarray}}

\newcommand{\lp}{\left(}
\newcommand{\rp}{\right)}

\newcommand{\nn}{\nonumber}
\newcommand{\appref}[1]{Appendix~\ref{app#1}}

\newcommand{\eqref}[1]{Eq.~(\ref{#1})}
\newcommand{\ct}[1]{~\cite{#1}}

\def\gev{\, {\rm GeV}}

\def\pb{\, {\rm pb}}

\newcommand{\sci}[2]{#1$\times$10$^{\text{#2}}$}

\newcommand{\tx}[1]{\mathrm{#1}}

\def\ev{\, {\rm eV}}

\usepackage{enumerate}
\include{epsf}

\begin{document}

\preprint{UH511-1198-12}
%\preprint{UTTG-06-11}
%\preprint{TCC-xxx-11}

\title{
%\PRE{\vspace*{1.3in}}
\textsc{Bremsstrahlung signatures of dark matter annihilation in the Sun}
\PRE{\vspace*{0.1in}}
}

\author{Keita Fukushima}
\affiliation{\mbox{Department of Physics \& Astronomy, University of
Hawai'i, Honolulu, HI 96822, USA}
%\PRE{\vspace*{.1in}}
}

\author{Yu Gao}
\affiliation{Department of Physics, University of Oregon, Eugene,
OR  97403, USA
%\PRE{\vspace*{.1in}}
}

\author{Jason Kumar}
\affiliation{\mbox{Department of Physics \& Astronomy, University of
Hawai'i, Honolulu, HI 96822, USA}
%\PRE{\vspace*{.1in}}
}

\author{Danny Marfatia\PRE{\vspace*{.1in}}}
\affiliation{\mbox{Department of Physics \& Astronomy, University of
Kansas, Lawrence, KS 66045, USA}
\PRE{\vspace*{.1in}}
}

%\date{\today}

%\pagestyle{headings}

\begin{abstract}
\PRE{\vspace*{.1in}}
The nonrelativistic annihilation of Majorana dark matter in the Sun to a pair of light fermions
is chirality-suppressed.
Annihilation to 3-body final states $\ell^+f^-V$,
where $V=W,Z,\gamma$, and $\ell$ and $f$ are light fermions (that may be the same), becomes dominant
since bremsstrahlung relaxes the chirality suppression.
We evaluate the neutrino spectra at the source, including spin and helicity dependent effects,
and assess the detectability of each significant bremsstrahlung channel at IceCube/DeepCore.
We also show how to combine the sensitivities  to the dark matter-nucleon scattering cross section
in individual channels, since typically several channels contribute in models.

\end{abstract}


\maketitle

%%%%%%%%%%%%%%%%%%%%%%%%%%%%%%%%%%%%%%%%%%%%%%%%%%%%%%%%%%

\section{Introduction}

Dark Matter (DM) particles $X$ can become captured and trapped at the center of the Sun and the Earth.
As the DM density grows over time,
the accumulated DM can annihilate and produce a neutrino flux that
is observable at a detector on Earth~\cite{bib:early_solar_dm_refs}.
The annihilation channels and the final state decay products
are determined by details of physics beyond the Standard Mode (SM).

While it is often assumed that 2-body annihilations dominate,
3-body final states can be the leading contribution in
models in which the DM candidate is a Majorana fermion.
Such candidates arise, for example in
supersymmetric models where the
lightest supersymmetric particle is a neutralino.
In such models, the cross section for dark matter annihilation
to light fermions is severely suppressed~\cite{bib:other_3body_papers}.

The wavefunction of an initial state consisting of a pair of identical fermions %the initial
must be totally antisymmetric, implying either $L=S=0$ or $L=S=1$, where $L$ and $S$ are the orbital angular momentum
and spin of the pair, respectively.
For the latter case, the annihilation matrix element
${\cal M}(XX \rightarrow f\bar f)$ is necessarily $p$-wave suppressed, and is thus proportional
to $v \ll 1$, where $v$ is the relative velocity of the DM particles.
The $s$-wave initial state is CP-odd and has zero total angular momentum; if CP-violating effects
are negligible, this state must annihilate to an $L=0$, $S=0$ final state.
As the final state fermions
$f \bar f$ emerge back-to-back, they must
possess the same helicity.
Since particles and antiparticles of the same handedness
arise from different Weyl spinors, $s$-wave $XX \rightarrow f \bar f$ annihilation
requires that the initial state couple to both the $f_L$ and $f_R$ spinors, {\it i.e.}, a mixed coupling to both L/R
chirality. (For further elaboration of these issues see the Appendix of Ref.~\cite{append}.)

While fermion mass readily provides L-R mixing, it leads to a matrix element suppressed by $m_f / m_X$.
If the mass term is the only source of helicity mixing,
the 2-body $XX \rightarrow f\bar f$ annihilation cross section is heavily suppressed.
However, a 3-body final state containing an additional vector boson (VB) can be
CP-even with vanishing total angular momentum, even if both fermions arise from the same Weyl spinor.
As a result, the 3-body annihilation cross section is not suppressed by $m_f^2 / m_X^2$,
and becomes significant despite the additional coupling factor ($\sim \alpha$).

Radiative electroweak corrections to DM annihilation were recently
considered for the gamma ray~\cite{Barger:2009xe,Bell:2011if,Barger:2011jg},
positron\ct{Bergstrom:2008gr,Kachelriess:2009zy} and antiproton~\cite{Kachelriess:2009zy,Garny:2011cj,Garny:2011ii} spectrum
of the annihilations. Recently, solar DM signals from electroweak bremsstrahlung
were investigated in Ref.~\cite{Bell:2012dk}.  In comparison, we consider
each significant annihilation channel separately; the corresponding event rates can be
summed using annihilation branching ratios
which depend on the details of a specific model.
We also consider DM annihilation to left-handed and right-handed fermions separately.
This is important, as the shape of the neutrino injection spectrum depends significantly on the helicity
of the fermions (and in particular on their decay spectra).
We also numerically propagate the neutrinos through the Sun and vacuum, with
oscillations, scattering and $\tau$-regeneration fully simulated.

In Section~\ref{sect:cross_section}, we describe the model we adopt and compute
the doubly differential 3-body annihilation cross sections.
The injection spectra are presented in Section~\ref{sect:injection}, and a description
of neutrino detection in Section~\ref{sect:result}.
In Section~\ref{sens},
we investigate the discovery potential of the annihilation channels individually and in combination at the IceCube/DeepCore
(IC/DC) detector. We conclude in Section~\ref{conc}.

\section{Cross sections}
\label{sect:cross_section}

Here we briefly discuss the annihilation cross section
in the case of $SU(2)$-singlet Majorana fermion dark matter $X$,
with a Lagrangian similar to that of Ref.~\cite{Ma:2000cc},
where $X$ couples to SM fermions $f$ through Yukawa terms,
\bea
L_{\tx{int}} =  y_L X P_L f \eta_L + y_L^* \bar f P_R X \eta_L^{*}  + y_R \bar X P_R f \eta_R^{}  + y_R^* \bar f P_L X \eta_R^{*}\,,
\label{eq:yukawa}
\eea
where the $y_{L,R}$ are Yukawa couplings, $\eta_L$ is a spin-0 $SU(2)$ doublet and $\eta_R$ is a spin-0 $SU(2)$ singlet.
Since $X$ is a gauge-singlet under the SM, the two-body annihilation $XX \rightarrow VV$ (where $V$ is a vector boson) does not occur
at tree level, even if kinematically allowed.
In general, one may also write a mixing term $\eta_L^* \eta_R + \eta_R^* \eta_L$, whose coefficient is proportional to the
Higgs vacuum expectation value. As L-R mixing lifts the suppression on $XX \rightarrow f\bar f$ (see,
for example,~\cite{Fukushima:2011df}), we restrict our attention to cases where such terms are negligible.

Setting $y_R =0$ without loss of generality, the leading contributions to the 2-body annihilation
cross section are given by~\cite{Bell:2011if}
\bea
v \sigma_{XX \to f \bar f} = \mathcal{O} \lp {m_f^2 \over m_{X}^2} \rp + \lp {y_L^4 \over 48 \pi m_{X}^2} {1+r^2 \over (1+r)^4}\rp v^2 +  \mathcal{O} (v^4)\,,
\eea
where $r = m_{\eta}^2 / m_X^2$. As expected, the first ($s$-wave) term is
suppressed by $m_f^2 / m_X^2 \ll1$, while the second ($p$-wave) term
is suppressed by $v^2$. Note that at freeze-out $v \sim 0.2$ is not negligible, and the speed of DM particles in the solar core
is much smaller than that in the galactic halo, $v \sim 10^{-3}$.

In comparison, with the emission of a VB, the cross section for
$X X \to f \bar f V$ can be expanded as\ct{Ciafaloni:2011sa}
\bea
v \sigma_{XX \to f \bar f V} \sim {g^2 \over 4 \pi m_{X}^2}
\left[   \mathcal{O} \lp {v^2 \over r^2} \rp + \mathcal{O} \lp {v^2 \over r^3} \rp+ \mathcal{O} \lp {1 \over r^4} \rp \right]\,.
\eea
As expected, the first two velocity-dependent ($p$-wave) terms are small compared to
the corresponding term in the 2-body cross-section because of the extra coupling factor ${g^2 / 4 \pi}$.
However, the third ($s$-wave) term is velocity-independent
and can be significant if $r$ is not too large.
In fact, for typical halo velocities, this 3-body annihilation process
dominates the 2-body process for $r < \mathcal{O}(10)$\ct{Ciafaloni:2011sa}.
Note that the $s$-wave term is not suppressed by the mass insertion
as the VB spin cancels the total spin of the two fermions, which
can have the same chirality.

%\section{Calculation of the Cross Sections}

To compute the 3-body annihilation cross section, in addition to the
interaction Lagrangian in Eq.~(\ref{eq:yukawa}), we need the matter-gauge boson interaction
vertices.  These can be derived from the Lagrangian kinetic terms (in standard notation),
\bea
{\cal L}_{\tx{D}} &=& \bar f \left\{ i \gamma^{\mu} \lp \partial_{\mu} - ie{1\over \sqrt{2} \sin^2 \theta_W}
(W_{\mu}^+ T_+ + W_{\mu}^- T_-)
% \right.\right.
 %\hspace{1.5cm}
 %\left.\left.
 -ie {T_3 - \sin^2 \theta_W Q \over \sin \theta_W \cos \theta_W} Z_{\mu}  - ieA_{\mu} Q \rp -m \right\} f\,,\nn
\eea
and
\bea
{\cal L}_{\tx{KG}} = &&
| \partial_{\mu} \eta |^2
-i \Bigg(
\partial_{\mu} \eta^* \left\{  e{1\over \sqrt{2} \sin^2 \theta_W} (W_{\mu}^+ T_+ + W_{\mu}^- T_-)
+ e{T_3 - \sin^2 \theta_W Q \over \sin \theta_W \cos \theta_W} Z_{\mu}  +eA_{\mu} Q \right\}  \eta
\nn \\ &&
-\left\{  e{1\over \sqrt{2} \sin^2 \theta_W} (W_{\mu}^+ T_+ + W_{\mu}^- T_-) + e{T_3 - \sin^2 \theta_W Q \over \sin \theta_W \cos \theta_W}
Z_{\mu}+eA_{\mu} Q \right\} \eta^*  \partial_{\mu} \eta
\Bigg)- m^2 | \eta |^2
\nn \\ &&
+ \left\{ e{1\over \sqrt{2} \sin^2 \theta_W} (W_{\mu}^+ T_+ + W_{\mu}^- T_-)  + e{T_3 - \sin^2 \theta_W Q \over \sin \theta_W \cos \theta_W}  Z_{\mu}
+ eA_{\mu} Q \right\}^2 | \eta |^2\,.
\eea
The matter-gauge boson interaction terms permit the tree-level process
$XX \rightarrow f \bar f V$ to proceed via $t$- or $u$-channel exchange of $\eta_{L,R}$, with the VB radiated
from either the external legs ({\it i.e.}, final state radiation), or from the virtual $\eta$ (referred to as internal bremsstrahlung).

\begin{figure}[h]
\centering
\subfigure[$M_A$]{
\includegraphics[width=2.0in]{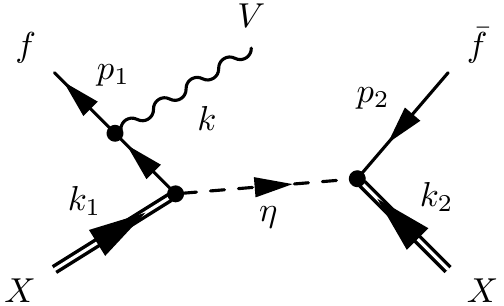}
\label{subfig1}
}
\subfigure[$M_B$]{
\includegraphics[width=2.0in]{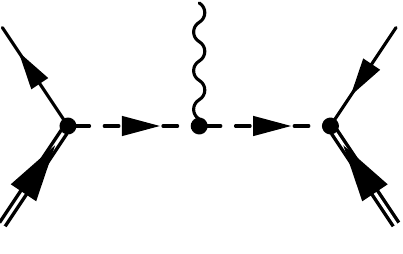}
\label{subfig2}
}
\subfigure[$M_C$]{
\includegraphics[width=2.0in]{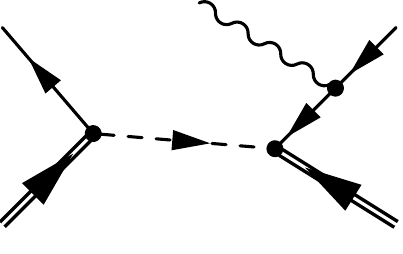}
\label{subfig3}
}
\subfigure[$M_A^{exc}$]{
\includegraphics[width=2.0in]{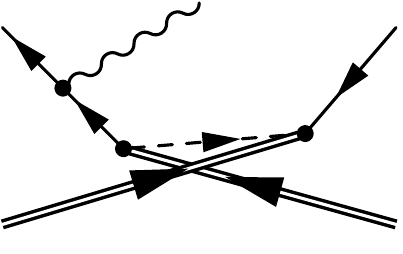}
\label{subfig4}
}
\subfigure[$M_B^{exc}$]{
\includegraphics[width=2.0in]{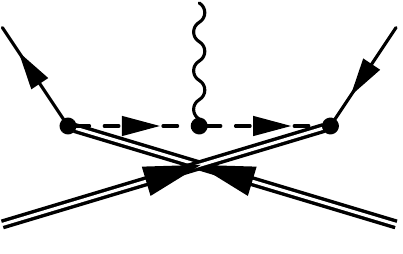}
\label{subfig5}
}
\subfigure[$M_C^{exc}$]{
\includegraphics[width=2.0in]{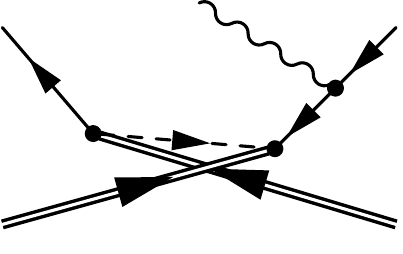}
\label{subfig6}
}
%\label{fig1}
\caption{Feynman diagrams for the 3-body annihilation process, $XX \rightarrow f\bar f V$.
The incoming DM particles $X$ have momenta $k_1$ and $k_2$. $f$, $\bar f$ are the outgoing
fermion and antifermion with momenta $p_1$ and $p_2$, respectively.  $\eta$ is the mediator particle,
and $V$ is a gauge boson ($\gamma$, $Z$, or $W$) with momentum $k$.
}
\label{fig:FeynmanDiagram}
\end{figure}

The total amplitude can be written as\ct{Ciafaloni:2011sa}
\bea
i {\cal M}  = i [(M_A + M^{exc}_A)+(M_B + M^{exc}_B)+(M_C + M^{exc}_C)]\,,
\eea
where the subscripts A-C refer to the diagrams in Fig.~\ref{fig:FeynmanDiagram}.
%The amplitudes in the case of $Z$-bremsstrahlung are also presented in Ref.\ct{Ciafaloni:2011sa}, which we reproduced.
%There is a relative minus between $t$- and $u$-channel amplitudes, arising from the interchange of fermionic operators.
We take the fermion masses to be negligible compared to the mass of the DM particle ($m_f / m_X \ll 1$).
%$k_1$ and $k_2$ are the two incoming DM momenta,
%and $p_1$ and $p_2$ are the momenta of $f$ and $\bar f$.
Using the {\it FeynCalc}~\cite{Mertig:1990an} package, we find that the squared matrix element (summed over polarizations and
averaged over initial spins) for annihilation to {\it e.g.}, $f_L \bar{f}_L Z$ is
\bea
|{\cal M}|^2_{XX \to f_L \bar{f}_L Z}
&=&
\frac{e^2 \left(1-2 \sin^2 \theta_W \right)^2 |y_L| ^4}{8  \sin^2 \theta_W \cos^2 \theta_W }
%\\ \nn
%&&\times
\frac{(4 (x_1 +x_2 -1)+\frac{m_Z ^2}{m_{X}  ^2}) \left(2 \left(x_1 ^2-2 x_1 +x_2 ^2-2 x_2 +2\right)-\frac{m_Z ^2}
{m_{X}  ^2}\right)}{m_{X}  ^2 (1-2 x_1 -r)^2 (1-2 x_2 -r)^2}\,,
\eea
where we use the notation of Ref.\ct{Barger:2011jg} and define
$x_1 =  E_{f_1} / m_X $, $x_2 =  E_{f_2} / m_X  $ and \mbox{$x_3 =  E_Z / m_X$}, in the static center of mass frame with $x_1 +x_2 +x_3 =2$.

We separately compute the
annihilation cross section to final states with any choice of fermion helicities. The differential cross sections,
\bea
v_{rel} {d\sigma \over dx_1 dx_2 }  = { |{\cal M}|^2  \over 128 \pi^3 }\,,
\eea
(where $v_{rel} = { v_1 - v_2 }$),  for all final state channels and helicities are given in \appref{2}; our results agree with those in
Refs.~\cite{Barger:2009xe,Barger:2011jg,Garny:2011ii,Ciafaloni:2011sa}.
We have checked that integrating the differential cross sections in the $m_Z \to 0$ limit, yields
the results in Refs.~\cite{Bell:2011if,Bergstrom:2008gr,Garny:2011cj,Ciafaloni:2011sa}.
An analytic expression for the total cross section is given in Ref.\ct{Bell:2011if}.

\section{Neutrino spectra}
\label{sect:injection}

In this section we discuss neutrino injection from
the leading annihilation channels with VB-bremsstrahlung.
We focus on the couplings of DM to leptons because these
are the most relevant to searches at neutrino detectors.  In particular, lepton couplings 
produce neutrinos directly as part of the 3-body final state, and
can provide a substantial contribution to the neutrino spectrum at 
high energy. However, lepton couplings do not contribute significantly to the DM
capture rate; although DM can scatter off electrons in the Sun, such 
collisions do not result in DM capture because the momentum transfer is
very small (since $m_e \ll m_X$).  Additional interactions between DM and light quarks  thus 
provide the dominant contribution to the capture rate. It is worth noting that, although 
DM-quark interactions can also induce annihilation, these are unlikely to produce energetic 
neutrinos.  These annihilation processes do not directly produce neutrinos in either the 2-body or 3-body 
final state, and the outgoing light 
quarks hadronize and stop before decaying, resulting in a very soft neutrino 
spectrum.  Henceforth, we simply assume that there are some additional DM-quark 
interactions responsible for DM capture in the Sun.

In general, the different annihilation
channels are not independent of each other, and their branching fractions are determined
by the couplings $y_{L,R}$ (assuming the $\eta_{L,R}$ states have degenerate mass).
We present the spectra for individual channels, however, in order to illustrate which
channels provide the hardest neutrino spectra.
Moreover, for models in which the degeneracy of the $\eta_{L,R}$ is broken, the total injection
spectrum can be found by summing the spectra of the individual channels after an appropriate
rescaling.

Lepton chirality plays
an important role for both the annihilation cross section and the shape of neutrino spectrum arising from
lepton decays. We investigate a scenario of flavor-independent lepton couplings, and a pure
third-generation coupling (100\% $\tau$) scenario.

We calculate the tree-level matrix element for the 4-body annihilation,
$XX\rightarrow f\bar{f}(V\rightarrow f\bar{f})$ for all choices of fermion
helicity.  Thus, contributions in which the gauge boson is produced
off-shell are included.  The SM quantum numbers of $\eta_{L,R}$ are determined by gauge-invariance,
and we assume that all scalar partners $\eta_{L,R}$ share a universal mass $m_\eta= \sqrt{r} m_X$.
Consequently, the matrix elements are entirely determined by $y_{L,R}$, $r$ and $m_X$.

The leading contributors to the neutrino spectrum are
\begin{itemize}
\item{Primary neutrinos produced directly from the annihilation
($XX \rightarrow \nu \bar \nu Z ,\, l^- \bar \nu W^+ ,\, l^+ \nu W^-$).}
\item{Secondary neutrinos produced from the decay of primary $W^\pm \rightarrow l^\pm \nu$,
$Z \rightarrow \nu\bar \nu$ produced in the annihilation
($XX \rightarrow \nu \bar \nu Z ,\, l^+ l^- Z, l^-\bar \nu  W^+ ,\, l^+ \nu W^-$).}
\item{Neutrinos from the decay of primary $\tau$/$\bar \tau$'s
produced in the annihilation process ($XX \rightarrow \bar \tau \tau (Z,\gamma),\, \bar \nu_\tau \tau W^+,\,
\bar \tau \nu_\tau W^-$).}
\end{itemize}

We study the following channels:

\hspace{1cm} (1) $XX\rightarrow \tau_L \bar \tau_L \gamma$

\hspace{1cm} (2) $XX\rightarrow \tau_R \bar \tau_R \gamma$

\hspace{1cm} (3) $XX\rightarrow \nu\bar{\nu}(Z\rightarrow\nu\bar{\nu})$\  for $\nu_e:\nu_\mu:\nu_\tau =1:1:1$

\hspace{1cm} (4) $XX\rightarrow \nu_\tau\bar{\nu}_\tau(Z\rightarrow\nu\bar{\nu})$  \hspace{1.cm} (100\%  $\nu_\tau\bar{\nu}_\tau Z$)

\hspace{1cm} (5) $XX\rightarrow \tau_L\bar{\tau}_L (Z\rightarrow\nu\bar{\nu})$

\hspace{1cm} (6) $XX\rightarrow \tau_R\bar{\tau}_R (Z\rightarrow\nu\bar{\nu})$

\hspace{1cm} (7) $XX\rightarrow \bar \tau_L \nu_\tau  (W^-\rightarrow l^-\bar{\nu}_l)\ +\ c.c$

\hspace{1cm} (8) $XX\rightarrow \bar{\l}_L \nu  (W^-\rightarrow l^-\bar{\nu}_l)\ +\ c.c$\ \ for $\nu_e:\nu_\mu:\nu_\tau =1:1:1$

\hspace{1cm} (9) $XX\rightarrow l^+l^-(Z\rightarrow \nu\bar{\nu},\bar \tau \tau )$, $l=e,\mu$

The subscripts $L$, $R$ refer to the helicity of the Weyl spinor ({\it e.g.}, $\bar \tau_L$ is the
antiparticle of a left-handed $\tau^-$, which is a right-handed $\tau^+$), and
`c.c.' denotes the CP conjugate process.
In addition to the three sources
of neutrinos described above, for channel (9) we also include tertiary neutrinos arising from the decay
of secondary $\tau$'s produced from primary $Z$ decay.
This is the only channel for which the tertiary contribution is significant.

\begin{figure}[t]
\includegraphics[scale=0.7]{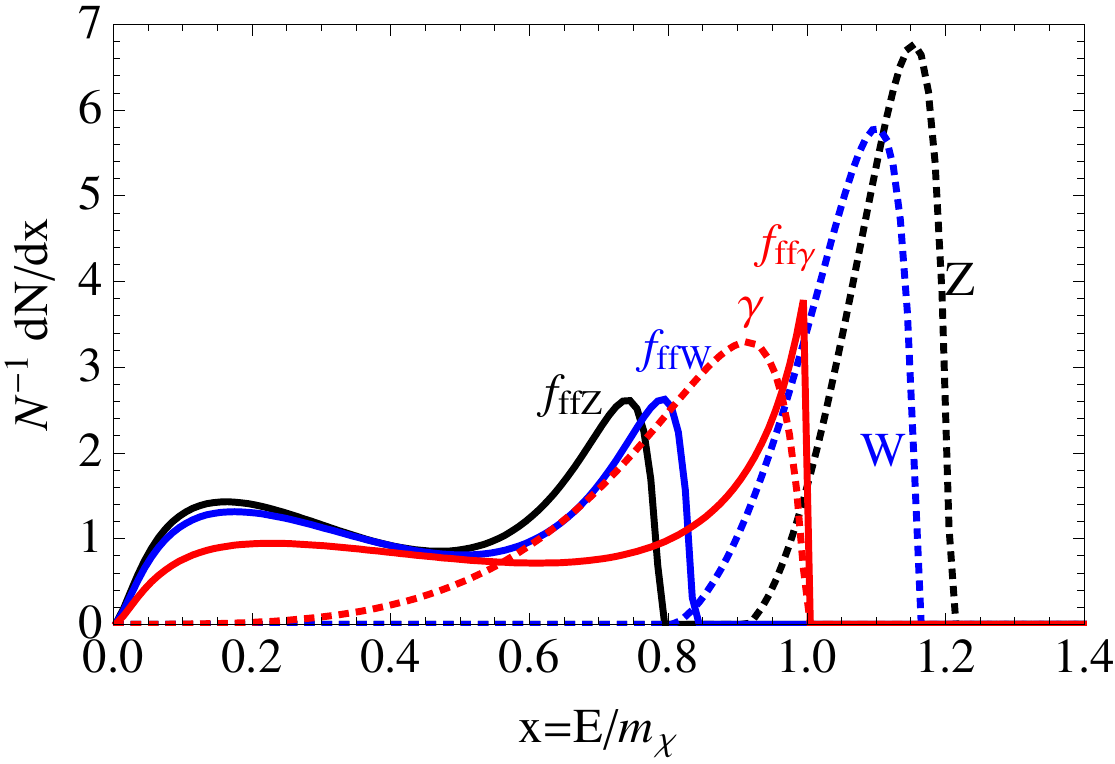}
\caption{
Normalized fermion spectra (solid) and VB spectra (dotted) for $XX \rightarrow f\bar{f}V$ for \mbox{$m_X=100$~GeV} and $m_\eta=105$~GeV
from Eq.~(\ref{eq:int_lvl1}).
%For a vector boson with momentum $k$, the energy is given by $E \equiv \sqrt{k^2 + m_V^2}$.
For light DM, off-shell VB emission is significant, resulting in 4-body spectra that deviate from these on-shell
calculations.
}
\label{fig:lvl1_injection}
\end{figure}

Since the dense solar medium readily absorbs electrons and muons,
channels (1-2) are the only channels with photon-bremsstrahlung that yield a significant neutrino flux.
In these channels neutrinos arise from $\tau$ decay, so the neutrino injection spectra are rather soft and are
dominated by the $\nu_\tau$ flavor. Also, since the photon does not decay,
the primary lepton and VB spectra can be obtained from the doubly-differential
3-body ($XX \rightarrow f\bar f V$) annihilation cross section,
\bea
\frac{d\sigma}{dx_1} &=& \int_{x_2^-}^{x_2^+}dx_2\frac{d\sigma}{dx_1dx_2}\,,\ \ \ \ \ \ \ \ \ \ \ \ \ \ 
%\nonumber\\
\frac{d\sigma}{dx_V}=\int_{x_1^-}^{x_1^+}dx_1\left.\frac{d\sigma}{dx_1dx_2}\right|_{x_2=2-x_1-x_V}\,,
\label{eq:int_lvl1}
\eea
where $x_1^{\pm}=\frac{1}{2}(2-x_V\pm\sqrt{x_V^2-r_V})$, $x_2^-= 1-x_1-r_V/4,\ x_2^+=1-r_V/(4(1-x_1))$,
and $r_V\equiv (m_V/m_X)^2$.  The lepton and VB spectra are plotted in Fig.~\ref{fig:lvl1_injection}
for $m_X = 100$~GeV and $m_\eta=105$~GeV.

Channels (3-6) involve $Z$-strahlung.  Primary $\nu\bar{\nu}$ are produced in channels (3-4), and provide the
dominant contribution to the neutrino spectrum.  Channels (5-6) lead to softer spectra because
the primary fermions are $\tau$ leptons. The case of $Z$-strahlung in which the primary fermions are $e / \mu$ yields a still
softer neutrino spectrum, and is treated separately in Channel (9).
Channels (7-8) involve $W$-bremsstrahlung (with either 100\% $\nu_\tau$ or flavor-independent couplings),
in which both primary and secondary neutrinos
and charged leptons appear. The coupling in $W$-bremsstrahlung is solely left-handed.  We only consider leptonic $W/Z$ decays,
as the hadronic decays suffer from absorption in the solar medium and produce considerably softer neutrinos.

\begin{figure}[t]
\includegraphics[scale=0.67]{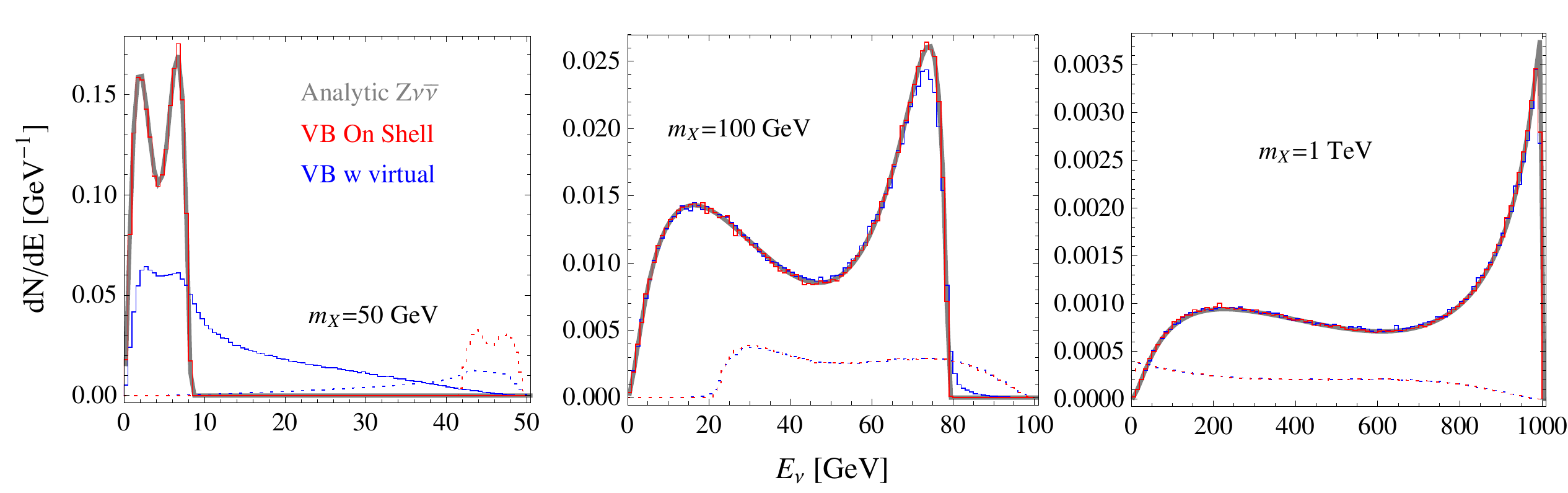}
\caption{Modification of the spectra from virtual VB production in the
$XX\rightarrow \nu\bar{\nu}Z$ channel for $m_X = 50$, 100 and 1000~GeV.
The primary (solid) and secondary (dotted) neutrino spectra are shown.
The `On Shell' and `w(ith) virtual' curves are generated using {\it Calchep}.
Spectra obtained from Eqs.~(\ref{eq:diff_ffZ}) and (\ref{eq:int_lvl1}) are also shown.
}
\label{fig:virtual}
\end{figure}

To incorporate virtual W/Z contributions in Channels (3-9), we use the numerical package {\it CalcHep}~\cite{Pukhov:1999gg}
to compute the primary and secondary
fermion spectra separately, and apply helicity-dependent decay to the $\tau$ leptons, if present.
In Fig.~\ref{fig:virtual}, we show the primary and secondary neutrino spectra from $XX \rightarrow \nu\bar \nu  Z$
for $m_X = 50, 100, 1000$~GeV, assuming either an on-shell VB, or allowing the VB to be off-shell.
For DM mass above 100~GeV, the virtual contribution becomes
subdominant and the primary spectra are reasonably well-described by
3-body cross sections.

\begin{figure}
\includegraphics[scale=0.62]{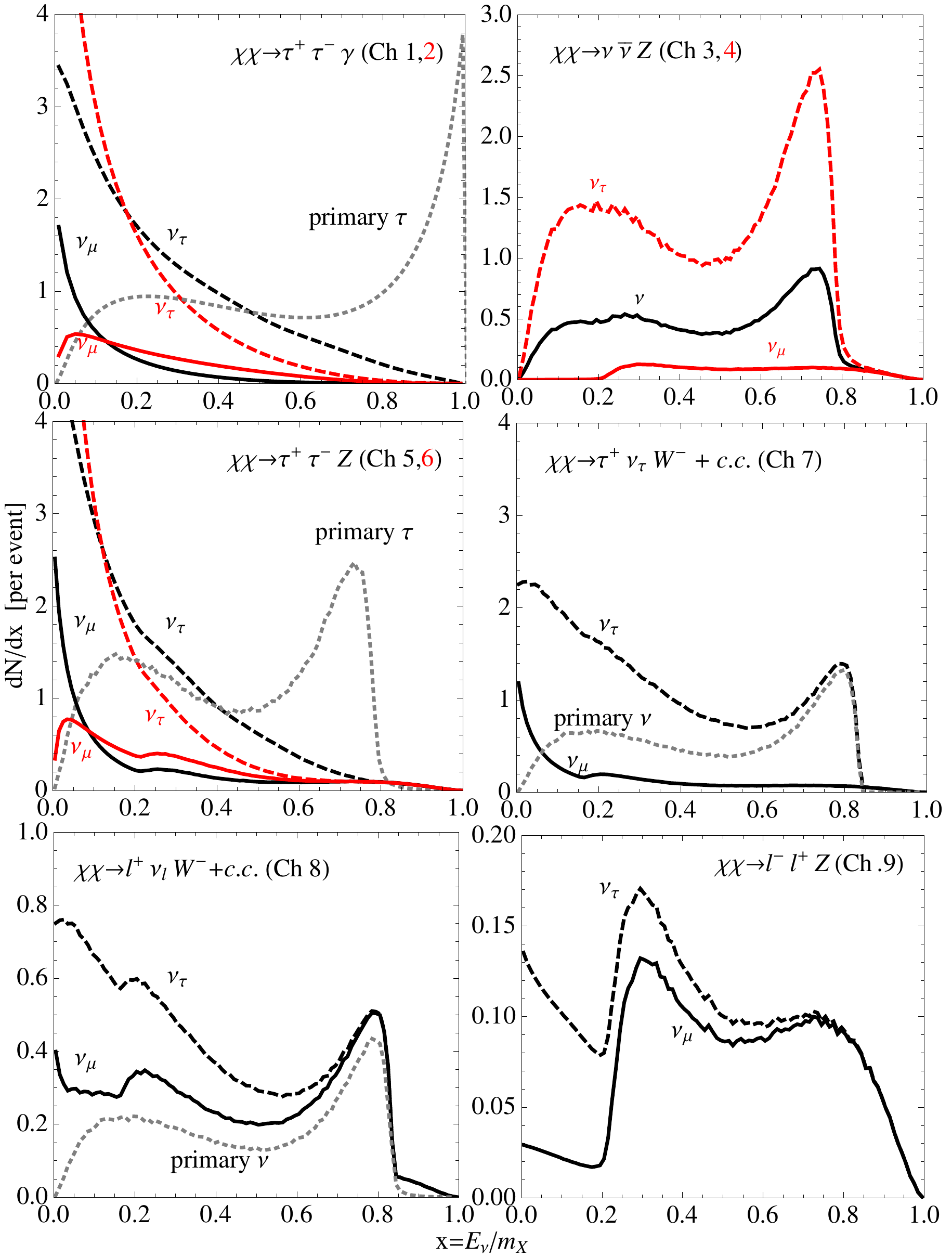}
\caption{
$\nu_\mu$ and $\nu_\tau$ injection spectra from each annihilation channel (as labeled),
including $\tau$ decay contributions, for $m_X = 100$~GeV and $m_\eta=105$~GeV.
The $\nu_e$ injection spectra are identical to that of $\nu_\mu$.
For comparison, the gray dotted curves show the primary lepton spectrum in each case.
The antineutrino injection spectra are identical.
}
\label{fig:lvl2_injection}
\end{figure}

\begin{figure}
\includegraphics[scale=0.62]{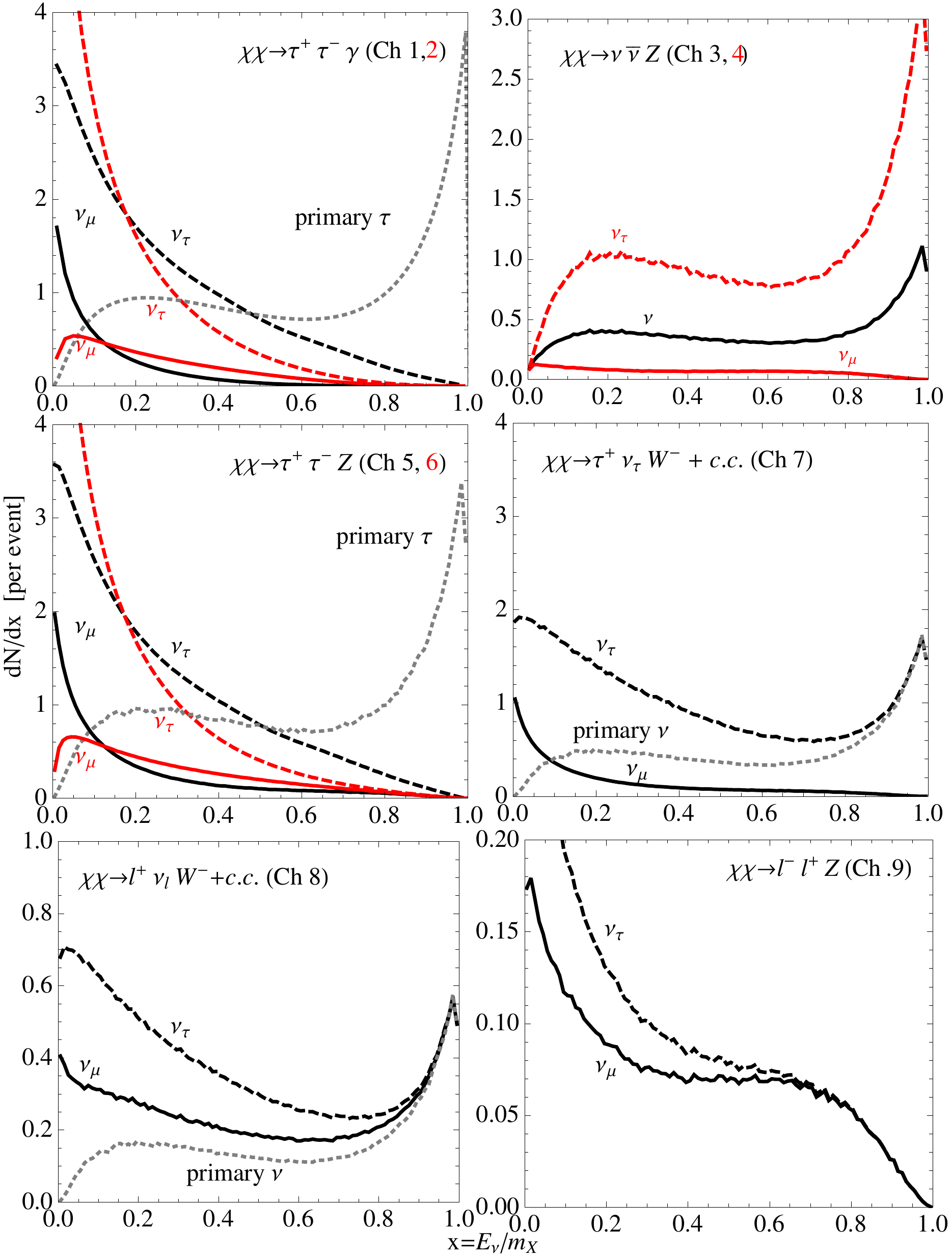}
\caption{
Similar to Fig.~\ref{fig:lvl2_injection}, for $m_X =1000$~GeV and
$m_{\eta}=1050$~GeV.
}
\label{fig:lvl2_injectionTeV}
\end{figure}

We display the neutrino injection spectrum from each channel for
$m_X=100~\gev$, $m_\eta=105~\gev$ (Fig.~\ref{fig:lvl2_injection}), and for
$m_X = 1000~\gev$, $m_\eta =1050~\gev$  (Fig.~\ref{fig:lvl2_injectionTeV}).
The antineutrino injection spectra are identical.

Although the 3-body differential cross section
for $XX \rightarrow \tau_R \bar \tau_R V$ can be obtained from
that of $XX \rightarrow \tau_L \bar \tau_L V$ by simply rescaling
by coupling factors (see Appendix~\ref{app2}), the resulting
neutrino spectra are quite different.  This can be seen by comparing the
spectra for \mbox{Channels (1)} and (2), and for Channels (5) and (6) in Figs.~\ref{fig:lvl2_injection} and
\ref{fig:lvl2_injectionTeV}.
The reason for this difference is that
the neutrino spectrum arising from the decay of a $\tau$ depends on its helicity;
the neutrino spectrum arising from a highly-boosted left-handed
$\tau$ differs markedly from that of a right-handed $\tau$.
For each channel, the fraction of left/right
handed $\tau$ is set by the couplings. $W$ decay produces 100\% left-handed $\tau$ while the left/right fraction from $Z$ decay
is 57\%/43\%.
See Ref.~\cite{Barger:2011em} for a description of our treatment.
Note that the channels dominated by $\tau$ decay lead to a rather soft
neutrino spectrum; its power-law shape is less distinctive compared to channels that yield primary neutrinos.

Due to the finite mass of the W/Z bosons, the neutrino spectra depend non-trivially on $m_X$.
Consider, for example, the $XX\rightarrow \nu\bar{\nu}Z$ channel:
if $m_X \sim m_Z / 2$, then the cross section for producing an on-shell $Z$ is suppressed by the phase space of the `primary' neutrinos
($E_\nu \sim m_X-m_{Z}/2$).  The primary neutrino can be much softer than $m_X$,
even below that of the secondary neutrinos, as shown in Fig.~\ref{fig:virtual}.
As can be seen from Fig.~\ref{fig:lvl2_injectionTeV}, for $m_X \gg m_Z$, the $\nu \bar \nu Z$
and $l \nu W$ channels are by far the hardest.

It is worth noting, however, that the finite mass of the the $W$ and $Z$ gauge bosons can also 
enhance the annihilation cross section to those channels.  The reason for this enhancement is 
that for some regions of phase space, the final state can only have vanishing total angular momentum if the gauge boson 
is helicity-$0$.  These contributions must vanish in the $m_V\to 0$ limit, implying that these terms in the
 squared matrix element scale as $m_V^2 / m_X^2$.

The mediator mass $m_\eta$ also has a noticeable impact on the shape of neutrino spectrum. In general a larger $m_\eta$ leads to a neutrino
energy distribution that is less peaked at the end-point; this effect could slightly enhance the signal rate, since at lower energy (yet
 above detector thresholds) neutrinos suffer less attenuation from scattering. However, for the values of $m_X$ under consideration,
varying $m_\eta$ yields only insignificant changes to the shape of the spectrum.
Moreover, the annihilation cross section is suppressed by $1/r^2$.  We use a low mediator mass
$r=1.1$ throughout.

\section{Neutrino Detection}
\label{sect:result}

After injection at the solar center, the neutrinos propagate through the Sun, and then through vacuum to Earth.
In doing so, they oscillate and undergo scattering.
We take the neutrino mixing parameters to be
$$
\delta m_{21}^2 = 8.1 \times 10^{-5}~\ev^2\,,
\hspace{0.3cm}
\delta m_{31}^2 = 2.2 \times 10^{-3}~\ev^2\,,
\hspace{0.3cm}
%\nonumber\\
\theta_{12} = 33.2^\circ\,,
\hspace{0.3cm}
\theta_{23} = 45^\circ\,,
\hspace{0.3cm}
\theta_{13} = 8.8^\circ\,, 
$$
where $\delta m^2_{ji}=m_j^2-m_i^2$ and we use the value of
$\theta_{13}$ recently measured by the Daya Bay experiment~\cite{An:2012eh}.
For details of our simulation of neutrino propagation (including oscillations,
tau-regeneration, and energy losses due to collisions),
and muon detection at IceCube/Deepcore, see
Refs.~\cite{Barger:2011em,Barger:2010ng}.

To assess the IC/DC sensitivity we choose a muon energy window, $E_{th}\leq E_\mu \leq m_X$,
where $E_{th}$ is the detector threshold.
%in which $N^{sig}/\sqrt{N^{atm}}$ is maximized.
%\be
%N^{sig}/\sqrt{N^{atm}}\,,
%\label{eq:single}
%\ee  
We target a 3$\sigma$ detection, $N^{sig}=3\sqrt{N^{atm}}$, where 
$N^{sig}$ and $N^{atm}$ are the number of signal and atmospheric background events, 
\bea
N^{sig} &=& \int dt\int_{E_{th}}^{m_X}{\cal V}(\theta(t), E_\mu) \frac{d\Phi^{sig}}{dE_\mu}dE_\mu\,,   \label{eq:up}\\
N^{atm} &=& {\pi \delta_\theta^2} \int dt\int_{E_{th}}^{m_X}{\cal V}(\theta(t), E_\mu) \frac{d\Phi(\theta(t))^{atm}}{dE_\mu d\Omega}dE_\mu\,.
\label{eq:con}
\eea
Here, $\Phi^{sig}$ and $\Phi^{atm}$ are the muon fluxes generated by the signal and atmospheric neutrinos incident on ice, respectively, and
${\cal V}$ is the effective dimension
of the IC/DC detector relevant to the event type.
For neutrinos arriving from the direction of the Sun, most charged leptons
scatter within a cone of half-angle,
\be
\delta_\theta = 20^\circ \sqrt{\frac{10~\text{GeV}}{m_X}}\,;
 \hspace{1cm}
\label{eq:cone_size}
\ee
see, {\it e.g.},~\cite{bib:jgk}.
Equation~(\ref{eq:cone_size}) gives the intrinsic scattering angle for incoming neutrinos with energy $\sim m_X$, which
is comparable to the IC detector's angular resolution, and which we use to estimate the atmospheric background.
%A realistic observation cone size is under adjustment.
% as it may also be affected by experimental uncertainties in the angular resolution.
%We use this formula as a qualitative estimate.
%Note that $\delta_\theta$  determines the amount of atmospheric
%background, the constraint on DM-annihilation rate scales
%linearly with $\delta_\theta$, and can be adjusted accordingly.
Note that since $N^{atm}$ scales quadratically with $\delta_\theta$, the
constraint on the DM annihilation rate scales linearly with $\delta_\theta$.

Besides the variation with zenith angle and event energy, ${\cal V}$ depends on whether the muon event is
{\it up-going} or {\it contained}~\cite{GonzalezGarcia:2009jc}. Up-going (contained) events refer
to upward going muon tracks that start outside (inside) the instrumented volume of the detector. A discussion of the effective detector dimensions
for these two types of events can be found in Refs.~\cite{Barger:2011em,GonzalezGarcia:2009jc}.
In this analysis we consider up-going muon events for IC and contained events for DC, and a
1~km$^3$ effective volume with a 70~GeV energy threshold as an estimate for IC contained events.
The effective IC area for up-going events falls rapidly below $E_{th}=$~60 GeV and we assume $E_{th}=$~10 GeV for the
DC subdetector.
Since observations track the trajectory
of the Sun, we use the zenith-angle dependent atmospheric $\nu_\mu$ flux measured by Super-Kamiokande~\cite{Honda:2006qj}.
The DM-induced muon flux has an energy spectrum that is determined by the DM
capture rate and the neutrino spectrum.
The time integral in Eqs.~(\ref{eq:up}-\ref{eq:con}) spans half a year for IC and a full
year for DC, which has 4$\pi$ angular coverage.
Descriptions of our calculations of up-going and contained event rates are provided in Refs.~\cite{Barger:2011em,Barger:2010ng}.

\section{Discovery Potential}
\label{sens}

\begin{figure}[t]
\includegraphics[scale=0.7]{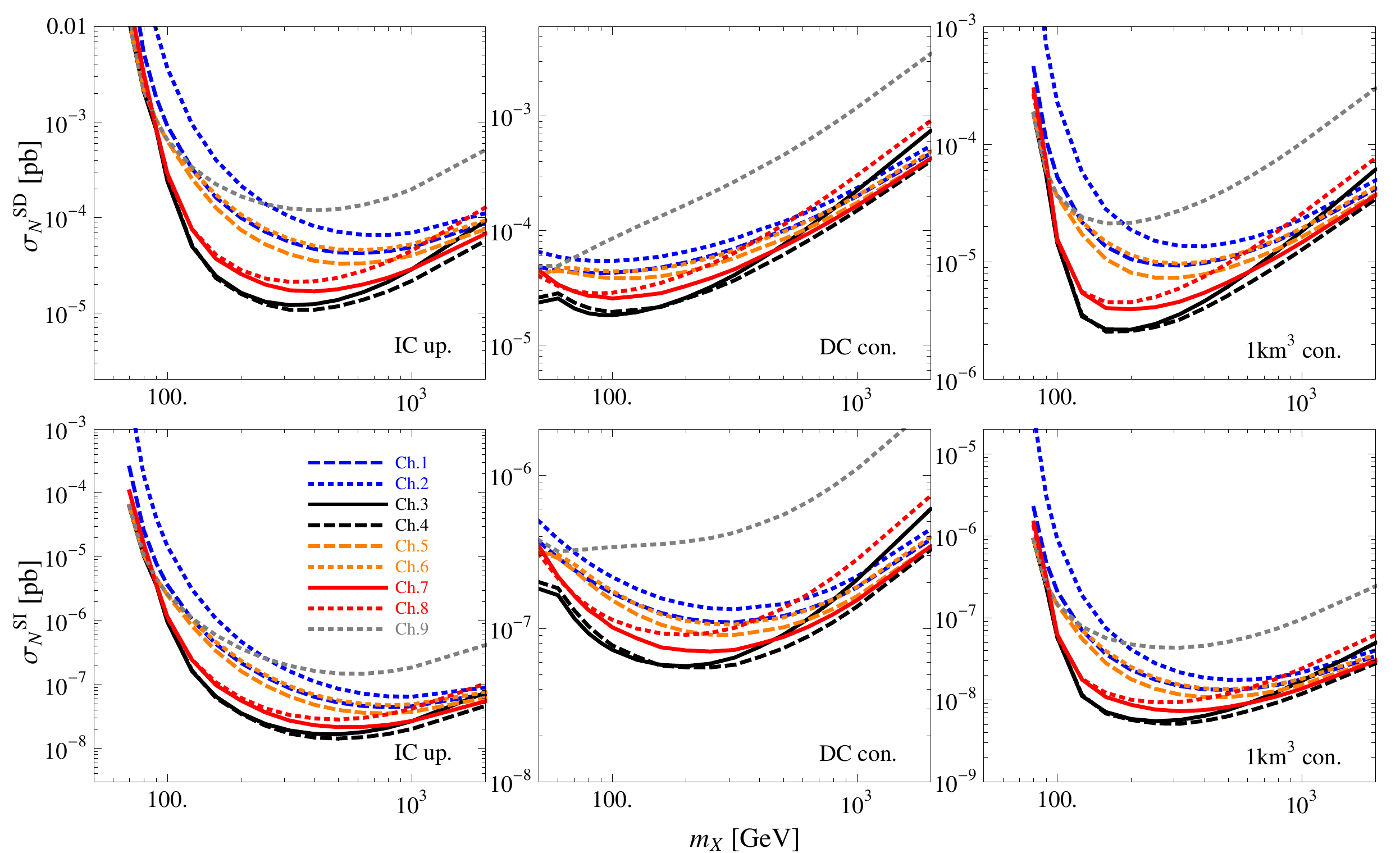}
\caption{The $3\sigma$ sensitivity to $\sigma_N^{SD}$ and $\sigma_N^{SI}$ of IC up-going, IC contained, and
DC contained events, with one year of data.  We assume that the DM capture and annihilation processes are in equilibrium, and set the
branching fraction to each channel equal to unity.}
\label{fig:sensitivity}
\end{figure}

\begin{figure}
\includegraphics[scale=0.9]{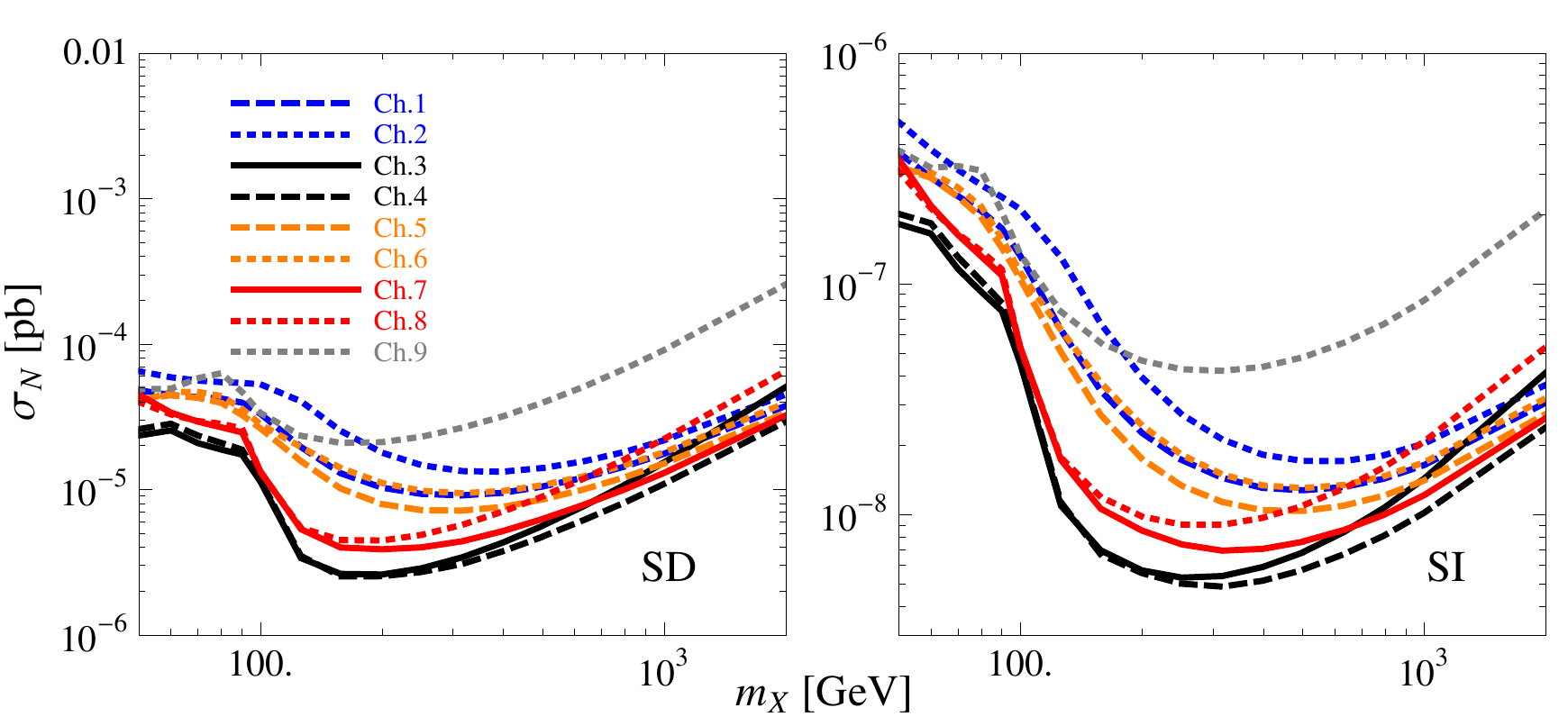}
\caption{The $3\sigma$ sensitivity to $\sigma_N^{SD}$ and $\sigma_N^{SI}$ of the combined
IC up-going, IC contained, and DC contained event samples, with one year of data.
In most of the mass range, contained events dominate
the signal rate.  DC dominates the sensitivity for $m_X$ below the
energy threshold of the less dense IC strings.
}
\label{fig:sensitivity_combined}
\end{figure}

IceCube's $3\sigma$ sensitivity to DM annihilation  can be determined by setting $N^{sig}= 3\sqrt{N^{atm}}$.
We assume that the DM capture and annihilation processes in the Sun are in equilibrium,\footnote{Whether dark matter capture and annihilation are in equilibrium is very model dependent. We pick the equilibrium case as a reference signal rate and present the rest of the calculation in as model-independent a manner as possible. Given the details of a specific model, the total annihilation rate (as a fraction of the capture rate) and the annihilation branching fractions can be determined, and our results can be translated into a bound on the particular model. \\ 
$\left. \ \ \right.$ Note that for $r=1.1$, $y_L=y_R=1$ and $m_X =100$~GeV, the total annihilation cross section to leptonic final states is 2.4~pb, and for 
$y_L=y_R =\sqrt{4\pi}$ and $m_X=1000$~GeV it is 0.8~pb. 
For annihilation cross sections of this order, and for scattering cross sections of the size to which IC/DC is sensitive, equilibrium will hold.
There can also be a contribution to the total annihilation cross section from quark final states (which provide only a subleading contribution to the neutrino spectrum).}
in which case a constraint on the annihilation rate can be directly
translated into a constraint on $\sigma_N \times BF_j$, the product of the DM-nucleon scattering cross section and the branching
fraction to the annihilation channel in question. We normalize the branching fraction to leptonic channels $BF({\rm leptons})$ to unity, so
 that $\sum_j BF_j=BF({\rm leptons})=1$. Therefore, the sensitivities to the DM-nucleon cross section 
 presented below should be divided by the actual value of $BF$(leptons).
In Table~\ref{tab:sensitivity_fixed_window_lite}, we present the sensitivity to each annihilation channel (using each of the 
three event samples) for $m_X= 50, 100$~GeV
and 1~TeV.
The $3\sigma$ sensitivity to the cross section for spin-independent (SI) and spin-dependent (SD) scattering are listed separately as
$\sigma_N^{SI}$ and $\sigma_N^{SD}$, respectively.
The dependence of the equilibrium annihilation rate on $\sigma_N$ is calculated using the method described in Ref.~\cite{bib:jgk}.
The corresponding background rates are also listed. In Fig.~\ref{fig:sensitivity}, we plot the $3\sigma$ sensitivity
to the DM-nucleon scattering cross section (setting the branching fraction to each channel equal to 1), under the
assumption that equilibrium holds between capture and annihilation in each individual channel. The combined sensitivity
of the three event samples obtained from
%\be
%\sigma_{tot} =
$\sqrt{\sum_{i=1}^3 \frac{\left(N^{sig}_i\right)^2}{N^{atm}_i}}=3$,
%\label{eq:combined}
%\ee
 is shown in Fig.~\ref{fig:sensitivity_combined} for each channel.
%Note that we have assumed Gaussian statistics; if an event sample contains only a small
%number of events, Poisson statistics would be more appropriate.
%However, since either the IC or DC contained events drive the statistical significance
%and these two have large number of events when they dominate, the assumption of Gaussian
%statistics does  not significantly impact the combined result.

%\begin{sidewaystable}
\begin{table}
\scriptsize
\begin{tabular}{|c|c|cc|cc|c|cc|c|cc|c|}
\hline
\multicolumn{2}{|c|}{} & \multicolumn{2}{|c|}{combined (pb)} &\multicolumn{3}{|c|}{IC up.}
&  \multicolumn{3}{|c|}{DC con.}  & \multicolumn{3}{|c|}{$1km^3$ con. ($E_\mu>70$~GeV)}\\
\hline
$m_X$(GeV),~$\delta_\theta$ & Ch.\# &$\sigma_N^{SD}$ & $\sigma_N^{SI}$ & $\sigma_N^{SD}$ & $\sigma_N^{SI}$ &  $\left.N^{atm}\right.^{}$
 &$\sigma_N^{SD}$ & $\sigma_N^{SI}$ & $\left.N^{atm}\right.^{}$
 &$\sigma_N^{SD}$ & $\sigma_N^{SI}$ & $\left.N^{atm}\right.^{}$\\
% &$\sigma^*_N (SD)$ & $\sigma^*_N (SI)$ & $E_{min}/E_{max}$ & $\left.N^{\mu~**}_{bkg}\right.^{}$\\
\hline
  &1 &\sci{4.8}{-5} &\sci{3.7}{-7} &  &   &  &\sci{4.8}{-5} &\sci{3.7}{-7}  & &  &  &    \\
  &2 &\sci{6.5}{-5} &\sci{5.0}{-7} &  &   &  &\sci{6.5}{-5} &\sci{5.0}{-7}  & &  &  &    \\
  &3 &\sci{2.4}{-5} &\sci{1.8}{-7} &  &   &  &\sci{2.4}{-5} &\sci{1.8}{-7}  & &  &  &    \\
$m_X=50$  &4 &\sci{2.6}{-5} &\sci{2.0}{-7} &  &   &  &\sci{2.6}{-5} &\sci{2.0}{-7}  & &  &  &    \\
$\delta_\theta=8.9^\circ$ &5 &\sci{4.2}{-5} &\sci{3.2}{-7} &  &   &  &\sci{4.2}{-5} &\sci{3.2}{-7}  &\sci{4.7}{2} &  &  &  \\
  &6 &\sci{4.0}{-5} &\sci{3.1}{-7} &  &    &  &\sci{4.0}{-5} &\sci{3.1}{-7}  & &  &  &    \\
  &7 &\sci{4.5}{-5} &\sci{3.5}{-7} &  &    &  &\sci{4.5}{-5} &\sci{3.5}{-7}  & &  &  &    \\
  &8 &\sci{4.9}{-5} &\sci{3.8}{-7} &  &    &  &\sci{4.9}{-5} &\sci{3.8}{-7}  & &  &  &   \\
  &9 &\sci{4.0}{-5} &\sci{3.1}{-7} &  &    &  &\sci{4.0}{-5} &\sci{3.1}{-7}  & &  &  &    \\
\hline
  &1 &\sci{3.3}{-5} &\sci{1.3}{-7} &\sci{9.1}{-4} &\sci{3.6}{-6} & &\sci{4.3}{-5} &\sci{1.7}{-7}  & &\sci{5.3}{-5} &\sci{2.1}{-7} & \\
  &2 &\sci{5.3}{-5} &\sci{2.1}{-7} &\sci{3.7}{-3} &\sci{1.5}{-5} & &\sci{5.4}{-5} &\sci{2.2}{-7}  & &\sci{2.3}{-4} &\sci{9.1}{-7} & \\
  &3 &\sci{1.1}{-5} &\sci{4.5}{-8} &\sci{2.4}{-4} &\sci{9.6}{-7} & &\sci{1.8}{-5} &\sci{7.2}{-8}  & &\sci{1.4}{-5} &\sci{5.8}{-8} & \\
$m_X=100$  &4 &\sci{1.2}{-5} &\sci{4.6}{-8} &\sci{2.5}{-4} &\sci{9.8}{-7} & &\sci{2.0}{-5} &\sci{7.8}{-8}  & &\sci{1.4}{-5} &\sci{5.7}{-8} & \\
$\delta_\theta=6.3^\circ$ &5 &\sci{2.6}{-5} &\sci{1.0}{-7} &\sci{6.4}{-4} &\sci{2.5}{-6} &24 &\sci{3.8}{-5} &\sci{1.5}{-7}  &\sci{2.7}{2} &\sci{3.6}{-5} &\sci{1.4}{-7} &\sci{2.8}{2} \\
  &6 &\sci{2.8}{-5} &\sci{1.1}{-7} &\sci{6.5}{-4} &\sci{2.6}{-6} & &\sci{4.4}{-5} &\sci{1.8}{-7}  & &\sci{3.6}{-5} &\sci{1.4}{-7} & \\
  &7 &\sci{1.3}{-5} &\sci{5.3}{-8} &\sci{2.9}{-4} &\sci{1.2}{-6} & &\sci{2.6}{-5} &\sci{1.0}{-7}  & &\sci{1.6}{-5} &\sci{6.2}{-8} & \\
  &8 &\sci{3.4}{-5} &\sci{1.3}{-7} &\sci{6.5}{-4} &\sci{2.6}{-6} & &\sci{8.5}{-5} &\sci{3.4}{-7}  & &\sci{3.7}{-5} &\sci{1.5}{-7} & \\
  &9 &\sci{1.3}{-5} &\sci{5.2}{-8} &\sci{2.8}{-4} &\sci{1.1}{-6} & &\sci{2.9}{-5} &\sci{1.1}{-7}  & &\sci{1.5}{-5} &\sci{5.9}{-8} & \\
\hline
  &1 &\sci{1.8}{-5} &\sci{1.6}{-8} &\sci{4.9}{-5} &\sci{4.5}{-8}  & &\sci{2.0}{-4} &\sci{1.9}{-7}  & &\sci{1.9}{-5} &\sci{1.8}{-8}  & \\
  &2 &\sci{2.2}{-5} &\sci{2.0}{-8} &\sci{6.9}{-5} &\sci{6.4}{-8}  & &\sci{2.3}{-4} &\sci{2.2}{-7}  & &\sci{2.3}{-5} &\sci{2.1}{-8}  & \\
  &3 &\sci{1.5}{-5} &\sci{1.4}{-8} &\sci{2.9}{-5} &\sci{2.7}{-8}  & &\sci{2.2}{-4} &\sci{2.1}{-7}  & &\sci{1.8}{-5} &\sci{1.7}{-8}  & \\
$m_X=1000$  &4 &\sci{1.1}{-5} &\sci{1.0}{-8} &\sci{2.2}{-5} &\sci{2.0}{-8}  & &\sci{1.5}{-4} &\sci{1.4}{-7}  & &\sci{1.3}{-5} &\sci{1.2}{-8}  & \\
$\delta_\theta=2^\circ$ &5 &\sci{1.5}{-5} &\sci{1.4}{-8} &\sci{4.0}{-5} &\sci{3.7}{-8}  &20 &\sci{1.7}{-4} &\sci{1.6}{-7}  &30 &\sci{1.6}{-5} &\sci{1.5}{-8}  &82 \\
  &6 &\sci{1.8}{-5} &\sci{1.7}{-8} &\sci{5.2}{-5} &\sci{4.9}{-8}  & &\sci{2.0}{-4} &\sci{1.9}{-7}  & &\sci{1.9}{-5} &\sci{1.8}{-8}  & \\
  &7 &\sci{1.3}{-5} &\sci{1.2}{-8} &\sci{2.9}{-5} &\sci{2.7}{-8}  & &\sci{1.6}{-4} &\sci{1.5}{-7}  & &\sci{1.5}{-5} &\sci{1.4}{-8}  & \\
  &8 &\sci{9.1}{-5} &\sci{8.5}{-8} &\sci{2.0}{-4} &\sci{1.8}{-7}  & &\sci{1.2}{-3} &\sci{1.1}{-6}  & &\sci{1.0}{-4} &\sci{9.6}{-8}  & \\
  &9 &\sci{2.2}{-5} &\sci{2.1}{-8} &\sci{4.5}{-5} &\sci{4.2}{-8}  & &\sci{3.0}{-4} &\sci{2.8}{-7}  & &\sci{2.6}{-5} &\sci{2.4}{-8}  & \\
\hline
%\multicolumn{10}{l}{$*\ \ $~Equilibrium $\sigma_N$ that gives $N_\text{sig}=3\cdot \sqrt{N_{\text{bkg}}}$ for $E_{min}<E<E_{max}$}\\
%\multicolumn{9}{l}{$*$~ IC rates are for half year.}\\
\end{tabular}
\normalsize
\caption{
The $3\sigma$ sensitivity to the spin-dependent (SD) and spin-independent (SI) DM-nucleon scattering cross sections for each annihilation
channel with one year of data. The number of atmospheric background events $N^{atm}$
in one year at DeepCore and 180~days at IceCube is also provided.
The muon energy window runs from the experimental threshold 
(10~GeV for DC, 60~GeV and 70~GeV for IC up-going and contained) to $m_X$.
}
\label{tab:sensitivity_fixed_window_lite}
%\end{sidewaystable}
\end{table}

%\newpage

%\newpage

For any particular model, there will be several annihilation channels and the total
event rate will be the sum of the contributions from each channel.  To compute the
sensitivity to the DM-nucleon cross section from a combination of channels, $\sigma_{N}^{comb}$,
we utilize the fact that signal significance scales linearly with
the cross section.
We denote by $\sigma_{N,i}^j$ the 3$\sigma$ sensitivity of IC/DC in annihilation Channel ($j$), 
using event sample $i$ (either DC contained events, IC up-going events, or IC 
contained events), as reported in Table~\ref{tab:sensitivity_fixed_window_lite}.  Denoting 
the branching fraction to Channel~($j$) by $BF_j$, we find the $3\sigma$ sensitivity for event 
sample $i$ from
\bea
\sigma_{N,i} &=& \left[\sum_j {BF_j \over \sigma_{N,i}^j } \right]^{-1}\,.
\eea
The $3\sigma$ combined sensitivity from a combination of channels and all three event types is then
\bea
\sigma_N^{comb} &=& \left[ \sum_{i=1}^3 {1 \over \sigma_{N,i}^2} \right]^{-{1\over 2}} \,.
\eea
In Table~\ref{tab:combinedbounds},
we show the $3\sigma$ sensitivity to $\sigma_N^{SD}$ and $\sigma_N^{SI}$ for
$m_X = 100, 1000$~GeV for three models: DM with couplings only to
left-handed third generation leptons ($y_R =0$), DM with couplings only to
to $\tau_R$ ($y_L =0$), and DM with equal couplings to left-handed and
right-handed third generation leptons ($y_L = y_R$).

\begin{table}[ht]
\begin{tabular}{|c|c|c|c|c|}
  \hline
  model & $\sigma_N^{SD}$ ($m_X = 100$~GeV) &$\sigma_N^{SD}$
($m_X = 1$~TeV)
  & $\sigma_N^{SI}$ ($m_X = 100$~GeV) &$\sigma_N^{SI}$ ($m_X =
1$~TeV) \\
  \hline
  $y_R =0$ & $1.4\times 10^{-5}~\pb$ & $1.3\times 10^{-5}~\pb$
  &$5.3 \times 10^{-8}~\pb$ & $1.2 \times 10^{-8}~\pb $  \\
  $y_L =0$ & $3.4\times 10^{-5}~\pb$ & $2.1\times 10^{-5}~\pb$
  & $1.4 \times 10^{-7}~\pb$ & $1.9 \times 10^{-8}~\pb$  \\
  $y_L = y_R$ & $1.4 \times 10^{-5}~\pb$& $1.4 \times 10^{-5}~\pb$
  & $5.6 \times 10^{-8}~\pb $  & $1.3 \times 10^{-8}~\pb $ \\
  \hline
\end{tabular}
\caption{The $3\sigma$ sensitivity to $\sigma_N^{SD}$ and $\sigma_N^{SI}$
(with 1 year of data) for three models with Yukawa couplings only to third generation
leptons: DM coupling only to left-handed
leptons ($y_R =0$), DM coupling only to $\tau_R$ ($y_L =0$), and DM
 with equal couplings to left- and right-handed leptons ($y_L = y_R$).
}
\label{tab:combinedbounds}
\end{table}

%\vspace{3.0in}

\section{Summary}
\label{conc}

If dark matter is a Majorana fermion, its annihilation in the Sun
to Standard Model fermions is both chirality and velocity-suppressed.
Then, the annihilation may primarily be through 3-body processes,
with the emission of a gauge boson.  The neutrino spectra from such channels can
differ dramatically from the spectra from $2\rightarrow 2$ processes.

Dark matter couplings to left-handed leptons necessarily open-up 3-body annihilation channels
in which neutrinos are produced directly.  Moreover, the branching fractions to these
channels are usually large.
The neutrino injection spectra are typically hard, providing for
interesting detection possibilities at neutrino detectors.

We considered a model in which $SU(2)$-singlet DM couples to
SM leptons (either left-handed or right-handed) via exchange
of a new scalar $\eta_{L,R}$.  For this model, we
calculated the 3-body differential annihilation cross sections
and the neutrino injection spectra
with a full treatment of the helicity correlations of the gauge boson and  $\tau$ lepton decays.
We determined the muon event rates at IceCube/DeepCore arising from each annihilation
channel, accounting for neutrino propagation effects, including
oscillation, scattering and regeneration.

%The chirality of the fermions that dark matter couples to affects both the annihilation rate and
%the shape of neutrino spectrum, via the helicity dependence of the $\tau$ decay spectra. The size of
%dark matter mass and that of the gauge boson determines whether virtual VB decays are significant,
%as at relatively low dark matter mass the space suppression from VB mass qualitatively alters the
%neutrino spectrum.

We calculated $3\sigma$ sensitivities of IC/DC to the DM-nucleon scattering
cross section for several 3-body channels.
The different channels are of varying utility in constraining dark matter models; channels with primary
neutrinos lead to the best sensitivity.

We then showed how to combine the sensitivities
in individual channels, to obtain the sensitivity
for a combination of channels as may arise in models.

%It is interesting to note that, for models in which dark matter which couples to
%first or second generation leptons, electroweak bremsstrahlung opens up 3-body annihilation channels
%in which hard primary neutrinos are produced.  Because muons and electrons typically stop in the sun before decaying,
%the neutrino spectrum produced from $XX \rightarrow \bar l l $ is very soft.  As a result, even if
%the cross-section for annihilation to fermion pairs is reasonably large, it is still difficult to distinguish
%the produced neutrino flux from atmospheric neutrino background.  The three-body annihilation channels with
%hard primary neutrinos can thus provide interesting detection signals for such models.

\vskip .2in
%\newpage
{\bf Acknowledgments.}
D.M. thanks the University of Hawaii for its hospitality during the initial stages of this work.
J.K. and D.M. thank the Center for Theoretical Underground Physics and Related Areas (CETUP* 2012) in South Dakota
for its support and hospitality during the completion of this work.
This research was supported in part by DOE
grants~DE-FG02-04ER41291, DE-FG02-04ER41308 and DE-FG02-96ER40969, and by
NSF grant PHY-0544278.

%\clearpage
%%%%%%%%%%%%%%%%%%%%%%%%%%%%%%%%%%%%%%%%%%%%%%%%%%%%%%%%%%
%\newpage
\appendix
\label{appendix}

\section{Vector boson emission cross sections}
\label{app:xsec}

As shown in Ref.~\cite{Barger:2011jg}, the analytic form of the differential cross section is independent of the emitted electroweak gauge boson.
Consequently, the set of differential and total cross sections for the various channels can be obtained from the following equations:

%We list the differential ($v_{rel} {d\sigma \over dx_1 dx_{2 \text{or} 3} } =   {|M|^2 \over 128 \pi ^3}$ ), total
%cross sections and conversion factors for different channels:% of $X X  \to f \bar f V $

%\scriptsize
\begin{eqnarray}
v_{rel} {d\sigma \over dx_1 dx_{2 \text{or} 3} } \Bigg|_{X X \to f_L \bar{f}_L Z}&=&
\frac{e^2 \left(1-2 \sin^2 \theta_W\right)^2 |y_L|^4}{1024 \pi ^3 \sin^2 \theta_W \cos^2 \theta_W}
\frac{(4 (x_1 +x_2 -1)+\frac{m_Z ^2}{m_{X}  ^2}) \left(2 \left(x_1 ^2-2 x_1 +x_2 ^2-2 x_2 +2\right)-\frac{m_Z ^2}{m_{X}  ^2}\right)}
{m_{X}  ^2 (1-2 x_1 -r)^2 (1-2 x_2 -r)^2}\,,
\nonumber
\end{eqnarray}
%\right\}
%\nonumber\\
%&\,&
%\\
%v_{rel} {d\sigma \over dx_1 dx_{2 \, \text{or}\,  3} } \Bigg|_{X X \to f_L \bar{f}_L \gamma}
%&=&
%\frac{e^2 |y_L|^4}{32 \pi ^3 }
%\left\{
%\frac{(x_1 +x_2 -1)  \left(x_1 ^2-2 x_1 +x_2 ^2-2 x_2 +2\right)}{m_{X}  ^2 (1-2 x_1 -r)^2 (1-2 x_2 -r)^2}
%\right\}
%\eea
%\normalsize

\label{app2}
%\scriptsize
\bea
\sigma_{XX \to f_R \bar{f}_R Z} &=&\frac{4 \sin^4 \theta_W}{\left(1-2 \sin^2 \theta_W\right)^2}  \sigma_{XX \to f_L \bar{f}_L Z} |_{y_L \to y_R}
\nn \\
\sigma_{X X \to \nu_L \bar{\nu}_L Z} &=&\frac{1}{\left(1-2 \sin^2 \theta_W\right)^2}  \sigma_{XX \to f_L \bar{f}_L Z}
\label{eq:diff_ffZ}
\nn \\
\sigma_{X X \to \nu_R \bar{\nu}_R Z} &=& 0
\nn \\
\sigma_{X X\to f_L \bar{\nu}_L W^{+ } }  = \sigma_{X X \to \bar{f }_L \nu_L W^{- }} &=& \frac{2 \cos ^2 \theta_W}{\left(1-2 \sin^2 \theta_W\right)^2}
\sigma_{X X \to f_L \bar{f}_L Z} |_{m_Z \to m_W}
\nn \\
\sigma_{X X \to f_R \bar{\nu}_R W^{+ } }  = \sigma_{X X \to \bar{f }_R \nu_R W^{- }} &=& 0
\nn\\
\sigma_{X X\to f_L \bar{f}_L \gamma}  &=& \frac{4 \sin^2 \theta_W \cos^2 \theta_W}{\left(1-2 \sin^2 \theta_W\right)^2}
\sigma_{X X \to f_L \bar{f}_L Z} |_{m_Z \to 0}.
\nn \\
\sigma_{X X\to f_R \bar{f}_R \gamma}  &=& \frac{ \cos^2 \theta_W }{ \sin^2 \theta_W}  \sigma_{X X \to f_R \bar{f}_R Z} |_{m_Z \to 0}
= \sigma_{X X\to f_L \bar{f}_L \gamma}  |_{y_L \to y_R}\,,
\nn
\eea
\normalsize

%%%%%%%%%%%%%%%%%%%%%%%%%%%%%%%%%%%%%%%%%%%%%%%%%%%%%%%%%%

%\label{app3}

where~\cite{Bell:2011if}

%%%%%%%%%%%%%%%%%%%%%%%%%%%%%%%%%%%%%%%%%%%%%%%%%%%%%%%%%%

%\label{app4}

%\scriptsize
\bea
&&
v_{rel}  \sigma_{X X \to f_L \bar{f}_L Z}
\nn \\ = &&
\frac{g^2 \left(1-2 \sin^2 \theta_W \right)^2 |y_L|^4}{1024 \pi ^3 \cos^2 \theta_W m_{X} ^2}
\Bigg\{
(r+1)
\Big[
{\pi^2 \over 6} - \ln ^2\left(\frac{2 m_{X} ^2 (r+1)}{4 m_{X} ^2 r-m_z ^2}\right)-2 \text{Li}_2\left(\frac{2 m_{X} ^2
(r+1)-m_z ^2}{4 m_{X} ^2 r-m_z ^2}\right)
\nn \\ &&
+2 \text{Li}_2\left(\frac{m_z ^2}{2 m_{X} ^2 (r+1)}\right) - \text{Li}_2\left(\frac{m_z ^2}{m_{X} ^2 (r+1)^2}\right) - 2
\text{Li}_2\left(\frac{m_z ^2[r-1]}{2 \left(m_{X} ^2[r+1]^2-m_z ^2\right)}\right)
\nn \\ &&
+2 \ln \left(\frac{4 m_{X} ^2 r-m_z ^2}{2 m_{X} ^2 (r-1)}\right) \ln \left(1-\frac{m_z ^2}{2 m_{X} ^2 (r+1)}\right)
+\ln \left(1-\frac{m_z ^2}{m_{X} ^2 (r+1)^2}\right) \ln \left(\frac{m_z ^2 (r-1)^2}{4 \left(m_{X} ^2 (r+1)^2-m_z ^2\right)}\right)
\Big]
\nn \\ &&
-\frac{m_z ^2 \left(4 m_{X} ^2 (r+1) (4 r+3)-(r-3) \left(m_z ^2-4 m_{X} ^2\right)\right)}{16 m_{X} ^4 (r+1)^2}
\nn \\ &&
+\frac{m_z ^2 \left(m_z ^4 (-(r-1))-2 m_z ^2 m_{X} ^2 (r+1) (r+3)+4 m_{X} ^4 (r+1)^4\right) }{4 m_{X} ^4 (r+1)^3 \left(m_{X} ^2 (r+1)^2
-m_z ^2\right)}\ln \left(\frac{m_z ^2}{4 m_{X} ^2}\right)
\nn \\ &&
+\frac{(r-1) \left(2 m_{X} ^2 (r+1)-m_z ^2\right) \left(-m_z ^6+2 m_z ^4 m_{X} ^2 (r (r+4)+1)-m_z ^2 m_{X} ^4 (r+1)^2 (3 r (r+6)+7)
+4 m_{X} ^6 (r+1)^4 (4 r+1)\right)}{4 m_{X} ^4 (r+1)^3 \left(4 m_{X} ^2 r-m_z ^2\right) \left(m_{X} ^2 (r+1)^2-m_z ^2\right)}
\nn \\ &&
\times \ln \left(\frac{2 m_{X} ^2 (r-1)}{2 m_{X} ^2 (r+1)-m_z ^2}\right)
+\frac{4 r+3}{r+1}\nn
\Bigg\}\,.
\eea

%\bea
%&&
%v_{rel}   \sigma_{X X \to f_L \bar{f}_L \gamma}
%\nn \\ = &&
%\frac{g^2 \sin^2 \theta_W |y_L|^2}{256 \pi^3 m_X^2}
%\left\{ (r+1) \left[ -2  \text{Li}_2\left(\frac{r+1}{2 r}\right)+\frac{1}{6} \pi ^2-\ln ^2\left(\frac{r+1}{2 r}\right)\right]
%+\frac{4 r+3}{r+1}+\frac{(r-1) (4 r+1) }{2 r} \ln \left(\frac{r-1}{r+1}\right)\right\}.
%\eea

\normalsize

%\vspace{3.0in}

%\vspace{0.7in}

%\clearpage

%%%%%%%%%%%%%%%%%%%%%%%%%%%%%%%%%%%%%%%%%%%%%%%%%%%%%

%%%%%%%%%%%%%%%%%%%%%%%%%%%%%%%%%%%%%%%%%%%%%%%%%%%%%%

\end{document}